\documentclass[prb,twocolumn,aps]{revtex4-2}
\usepackage{graphicx,amsmath,xcolor}
\usepackage{braket,hyperref}
\begin{document}

    \title{Dissipation and gate timing errors in SWAP operations of qubits}

    \author{Nathan L. Foulk}
    \author{Robert E. Throckmorton}
    \author{S. Das Sarma}
    \affiliation{Condensed Matter Theory Center and Joint Quantum Institute, Department of Physics, University of Maryland, College Park, Maryland 20742-4111 USA}

    \begin{abstract}
        We examine how dissipation and gate timing errors affect the fidelity of a sequence of SWAP gates on a chain of interacting qubits in comparison to noise in the interqubit interaction.
        Although interqubit interaction noise and gate timing errors are always present in any qubit platform, dissipation is a special case that can arise in multivalley semiconductor spin qubit systems, such as Si-based qubits, where dissipation may be used as a general model for valley leakage. 
        In our Hamiltonian, each qubit is coupled via Heisenberg exchange to every other qubit in the chain, with the strength of the exchange interaction decreasing exponentially with distance between the qubits.
        Dissipation is modeled through the term $-i\gamma\mathbf{1}$ in the Hamiltonian, and $\gamma$ is chosen so as to be consistent with the experimentally observed intervalley tunneling in Si.
        We show that randomness in the dissipation parameter should have little to no effect on the SWAP gate fidelity in the currently fabricated Si circuits.
        We introduce quasistatic noise in the interqubit interaction and random gate timing error and average the fidelities over 10,000 realizations for each set of parameters. 
        The fidelities are then plotted against $J_\text{SWAP}$, the strength of the exchange coupling corresponding to the SWAP gate.
        We find that dissipation decreases the fidelity of the SWAP operation---though the effect is small compared to that of the known  noise in the interqubit interaction---and that gate timing error creates an effective optimal value of $J_\text{SWAP}$, beyond which infidelity begins to increase.
    \end{abstract}
    \maketitle

    \section{Introduction}

        In quantum computation, a SWAP gate is required to increase connectivity beyond nearest neighbors, i.e., to transport quantum information across the system.
        The SWAP gate is performed by allowing two qubits to experience a greater coupling strength $J_\text{SWAP}$.
        This pulse lasts for time $\tau= \frac{\pi}{4J_\text{SWAP}}$ ($\hbar = 1$). 
        One can also perform its root $\sqrt{\text{SWAP}}$ by halving the gate time to entangle qubits.
        This makes SWAP, along with single-qubit gates, sufficient to achieve universal quantum computation \cite{Loss1998}. 
        The SWAP gate is therefore an essential element of quantum computation, and thus studying and reducing the error in the SWAP gate is critical to the eventual construction of a quantum computer.
        
        Quantum dot semiconductor spin qubits are promising candidates to form the basis of a quantum computer \cite{Loss1998}. 
        Such qubits benefit from long coherence times and relatively straightforward scalability.
        However, no one to date has demonstrated two-qubit gates on spin qubits with the fidelity required for the execution of quantum error-correcting codes, which are the basis of robust quantum computation.
        This level of fidelity is often referred to as the error-correction threshold, and is at least 99.9\%.
        Although the error-correction threshold has not yet been achieved for both single- and two-qubit gates, recently several experimental groups have come very close \cite{Huang2019,Mills2021,Noiri2021,Xue2021}.
        In order to achieve this error-correction threshold, recent experimental work \cite{Huang2019} suggests that fast gate times, much shorter than any decoherence time, are key.
        Gates based on the underlying Heisenberg exchange interqubit coupling in semiconductor spin qubits are very fast, since the exchange coupling can be controlled electrostatically. 
        Such gates include the SWAP gate for single-spin qubits, while multi-spin qubits have been designed so that all gates, single- and two-qubit, are performed simply by tuning the Heisenberg exchange coupling \cite{Medford2013,Eng2015}.
        
        Silicon has been put forward as an ideal material for semiconductor QD qubits for several reasons. 
        First, coherence times in Si are significantly longer than in GaAs systems, where decoherence due to Overhauser (nuclear spin) noise is unavoidable.
        Silicon's coherence times can be made even longer through isotopic purification, in which ${}^{29}$Si, silicon's only stable magnetic isotope, is removed in favor of the stable spin-0 isotopes ${}^{28}$Si and ${}^{30}$Si.
        GaAs systems cannot be purified in such a manner because there are no stable spin-0 isotopes of gallium nor arsenic.
        Spin qubits based on isotropically purified Si (so as to reduce nuclear spin induced decoherence) are thus particularly well-suited for high fidelity gate operations because the coherence time is usually long and the electric field induced gate operations are short, providing a large dimensionless fidelity efficiency ratio of coherence time to gating time.
        Second, Si-based qubits can be more easily integrated into the existing semiconductor industry, which is already capable of creating complex silicon processors made up of a very large number of transistors, or classical bits, thus making scaling up an easier task in Si-based systems.
        Therefore, a Si-quantum computer platform can, in principle, host many million, perhaps even billions, of qubits in a small chip similar to the existing CMOS-based integrated circuits.
        
        One well-known source of SWAP gate error is charge noise, which is ubiquitous and is thought to be the primary decohering mechanism in Si-based spin qubits since Overhauser noise can be effectively suppressed in Si.
        For semiconductor spin qubits, charge noise comes from charged defects at the interface and from noise in the applied gate voltages, which are always present in the environment. 
        This affects the exchange coupling directly, since the exchange coupling is fundamentally based on the interelectron Coulomb interaction between neighboring QDs \cite{Hu2006}.
        Much work is being done to mitigate the effects of charge noise, including composite pulses \cite{Wang2014} and symmetric gate operation \cite{Reed2016}.

        Another disadvantage that Si-based qubits in particular suffer from when compared to other semiconductor (e.g., GaAs) QD spin qubits is valley degeneracy.
        Bulk Si has six degenerate ellipsoidal valleys along the symmetry axes in the conduction band. 
        This sixfold degeneracy becomes twofold in the size-confined Si $[100]$ surface where the typical QDs hosting the spin qubit reside, with the other four valleys getting lifted in energy by virtue of the mass anisotropy. 
        This twofold ground-state degeneracy is often found to be lifted by a relatively small valley splitting $\Delta$, which depends on the interface sharpness and other factors that are often random and difficult to control because they depend on the microscopic details of the interface.
        For large $\Delta$, the system behaves as a spin qubit since the valley degree of freedom is effectively frozen as long as the operational temperature is much smaller than the valley splitting.
        When $\Delta$ is small, however, valleys create an avenue of decoherence during gate operation, known as valley leakage.
        This valley leakage problem, arising from any intervalley electron hopping, is independent of the problem associated with the higher valley thermal occupancy, and as such, may not necessarily be suppressed by making the temperature much lower than the valley splitting.
        Valley degeneracies also hinder the Pauli spin blockade, which is crucial to qubit initialization and readout \cite{Zwanenburg2013}.
        Although much work has been done \cite{Saraiva2011,Buterakos2021,Chen2021,Liu2021} to characterize valley leakage and develop methods to minimize information loss, some loss is inevitable because of possible intervalley coupling.
        
        One of the main goals of this work is to extend previous work on the fidelity of a sequence of SWAP gates in a chain of single-spin qubits by accounting for dissipation.
        Previous work by some of the authors \cite{Throckmorton2020} simulated the fidelity of a sequences of SWAP gates in the presence of charge noise.  
        Our work accounts for not only charge noise, but also dissipation.  
        The charge noise, as before, is assumed to be quasistatic, i.e., we model it as disorder in the interqubit interaction, which is drawn from a Gaussian distribution.  
        The dissipation term in the Hamiltonian $-i\gamma\mathbf{1}$ can serve as a phenomenological model of valley leakage in Si-based spin qubits and, in that context, would arise from ``tracing out'' one of the valleys.  
        As pointed out earlier, the hopping of an electron from one valley into another hinders measurement of that electron's state, and thus this other valley acts effectively as a bath.  
        Although there is no rigorous proof that such a dissipative term is a precise, quantitatively accurate description of valley leakage, we nonetheless believe that such a term provides a reasonable, physically-motivated, phenomenological model for valley leakage.
        However, our work is also a general, phenomenological, single parameter (i.e. $\gamma$) model for describing dissipation in the context of qubit operations.
        We consider randomness in this dissipation term, which represents the fact that actual semiconductor surfaces will not be perfectly smooth, causing variation in the intervalley coupling, by drawing the value of $\gamma$ from a Gaussian distribution that is truncated to nonnegative values.  
        Unlike charge noise, this random valley dissipation is truly static; the only similarity is in our mathematical treatment of these two effects.  
        We find that this variation in $\gamma$, however, has no readily discernible effect on the fidelity of the SWAP operations.  
        There is, however, some effect from changes in the ``average'' value $\gamma_0$, though the effect is small compared to that of charge noise for realistic values of $\gamma_0$.

        As more experimental progress is made, we expect gate times to become shorter and for gate timing error to become increasingly relevant, since precise timing, by definition, is impossible, and quantum computing necessitates very low error thresholds.
        Gate timing error refers to unintentional experimental errors in the gate duration $\tau$.
        This slight randomness in the SWAP time is inevitable, since the experimental controls cannot be absolutely precise.  
        We therefore include the effects of random gate timing error in our simulations as well.  
        We find that this timing error results in an effective ``optimal'' value of $J_\text{SWAP}$ that minimizes the infidelity of the SWAP gate sequence.

        The rest of the paper is organized as follows.
        We first introduce the model qubit Hamiltonian on which we base our work in Section II. 
        In Section III, we analytically examine the effects of dissipation and randomness in the dissipation parameter $\gamma$ and present our numerical results on SWAP fidelities in the presence of charge noise, dissipation, and gate timing error.
        Finally, we present our conclusions in Section IV.

    \section{Model}
        We model the system as a 1D spin chain using the following Hamiltonian,
        \begin{equation}
        \label{ham}
        H =  \sum_{k=1}^{L-1} \sum_{n=1}^{L-k} J^k_n \boldsymbol{\sigma}_n \cdot \boldsymbol{\sigma}_{n+k} - i \gamma \mathbf{1},
        \end{equation}
        where $L$ is the length of the spin chain and $J_n^k$ refers to the exchange coupling strength between the $n^\text{th}$ and $(n+k)^\text{th}$ qubit. 
        For example, $k=1$ corresponds to nearest-neighbor coupling, $k = 2$ corresponds to next-nearest-neighbor coupling, and so on. 

        The non-Hermitian term $-i \gamma \mathbf{1}$ is a dissipative term, making the time evolution operator nonunitary. 
        This term can represent the information loss from valley leakage arising from intervalley electron tunneling which effectively ``removes'' the electron from the qubit subspace, thus acting as a dissipation process.
        We use $\gamma$ as a phenomenological parameter to characterize the valley leakage.
        Of course, the parameter $\gamma$ must be positive in order to properly simulate leakage, or else the probability would increase exponentially with time.
        
        It is worth emphasizing that nothing in our modelling of dissipation is unique to Si-based spin qubits. 
        Indeed our approach is very general, and applies equally well to any two-level system that exhibits dissipative dynamics. 
        Si-based qubits are simply a prime candidate of such a system that is also currently relevant and experimentally feasible.
        Accordingly, we estimate the dissipation term using experimental values measured in a silicon triple quantum dot \cite{Petta2021}. 
        
        Very little is understood regarding why such intervalley leakage occurs, since the two valley states would be expected to be orthogonal.
        For this reason, the best we can do is a one parameter phenomenology.
        Valley leakage is the coherent evolution in a Hilbert space that is larger than our computational subspace, which is precisely what happens whenever decoherence or dissipation occurs, and therefore our choice of modeling this process as dissipation is justified. 
        In this case, the other valley states act as the ``bath'', and our computational subspace is the ``system''.
        
        Despite our poor theoretical understanding of how valley leakage occurs, we are fortunately able to estimate $\gamma$ using recent experimental results \cite{Petta2021}.  
        We estimate $\gamma$ to be no greater than one tenth the value of the nearest-neighbor coupling strength $J^1$, perhaps even much smaller.

        The previous work we are extending \cite{Throckmorton2020} simulated the effects of charge noise, but only included interactions between nearest and next-nearest neighbors. 
        In addition to the new physics of dissipation as described above, we extend the range of the exchange interactions to the whole length of the chain, with the interaction strength decreasing exponentially such that $J_n^{k+1}/J_n^k \equiv \beta < 1$.
        For example, if $\beta = 0.01$, then $J_n^3 = J_n^2/100 = J_n^1/10000$.
        In this way, each spin in the chain is coupled to every other spin in the chain. 
        This is consistent with the energetics of interdot exchange couplings, which ultimately arise from the electron wavefunction  overlaps between the QDs.  

    \section{Calculations}
        We now turn our attention to calculating the fidelity of our sequence of SWAP gates.  
        We define fidelity as
        \begin{equation}
            F \equiv | \langle \Psi_0 | R^\dagger U | \Psi_0 \rangle |^2,
        \end{equation}
        where $R$ is an operator which performs the SWAP gate sequence with perfect fidelity, $U$ represents the actual SWAP gate sequence with errors, and $\ket{\Psi_0}$ is the initial state of the system.
        In our case, the SWAP sequence is intended to transport a spin state from one side of the spin chain to the other, and then back again.
        Therefore, if the SWAP gates were ideal, the we expect our sequence of gates to be equivalent to the identity, so that $R = 1$.
    \subsection{Noise in the dissipative term \label{analytic}}
        Our simple dissipative model allows for the effects of $\gamma$ on fidelity to be calculated analytically, as it simply adds $-i\gamma$ to all of the energy eigenvalues without changing the corresponding eigenstates.  
        We find that
        \begin{equation}
        F = F_0 e^{- 2 \gamma t},
        \end{equation}
        where $F_0$ is the fidelity without any dissipation.

        We now show that the effects of randomness in the valley term on the fidelity are negligible.
        Assuming that noise in $\gamma$ is independent from other sources of disorder, we can isolate it so that
        \begin{equation}
            \langle F \rangle = F_0 \langle e^{-2 \gamma t} \rangle.
        \end{equation}
        
        We then introduce randomness in the dissipation term $\gamma$, taken from a Gaussian distribution with standard deviation $\sigma_\gamma$, truncated to positive values of $\gamma$.  
        We require that $\gamma>0$ so that the probability does not grow arbitrarily large with time.  
        We can calculate the average, $\langle e^{-2\gamma t}\rangle$, analytically:
        \begin{align}
            \langle e^{-2\gamma t} \rangle &= \frac{1}{\alpha\sigma_\gamma}\sqrt{\frac{2}{\pi}}\int_0^\infty e^{-2 \gamma t} e^{-(\gamma - \gamma_0)^2/2 \sigma_\gamma^2}\,d\gamma \\
             &=  e^{-2 \gamma_0 t}  e^{2 \sigma_\gamma^2 t^2 } \frac{1}{\alpha}\left(1 + \text{erf}\left[\frac{\gamma_0/\sigma_\gamma   - 2 \sigma_\gamma t}{\sqrt{2}}\right]\right)
        \end{align}
        where $\text{erf}(x)$ is the Gaussian error function and
        \begin{equation}
            \alpha = 1 + \text{erf}\left (\frac{\gamma_0}{\sigma_\gamma\sqrt{2}}\right ).
        \end{equation}
        
\begin{figure} 
            \centering
            \includegraphics[scale=0.3]{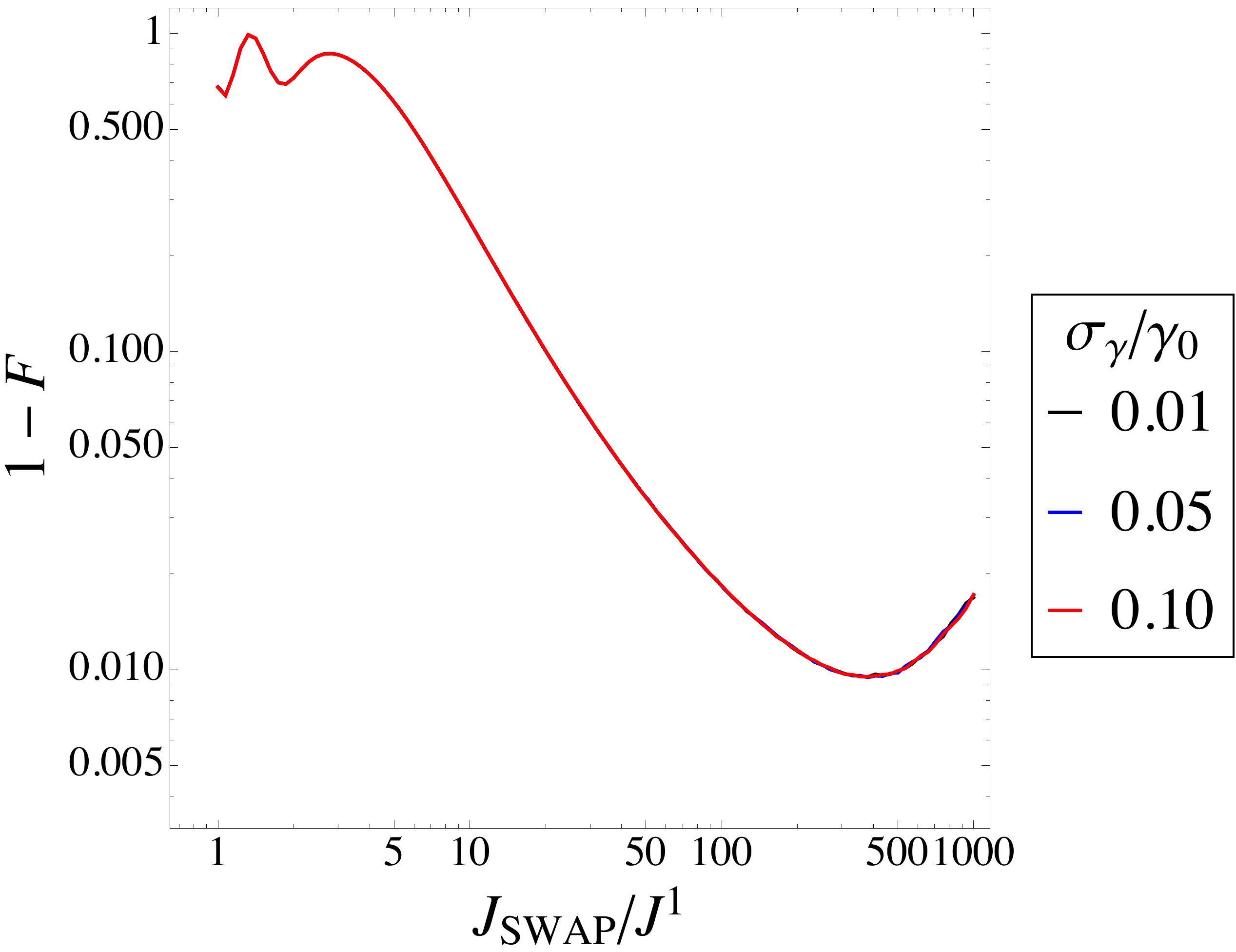}
            \caption{\label{fig:sig_gamma} The infidelity $1-F$ as a function of $J_\text{SWAP}/J^1$ for several values of $\sigma_\gamma$ is shown. 
            We see that increasing the parameter $\sigma_\gamma$ has little to no effect on the SWAP fidelity. 
            For this case, $\gamma_0/J^1 =  0.10$.}
        \end{figure}

        For reasonable values of our parameters, such  that $\sigma_\gamma/\gamma_0 < 0.1$ and $t \ll 1$, we see that
        \begin{equation}
            \langle e^{-2\gamma t} \rangle \approx e^{-2 \gamma_0 t},
        \end{equation}
        which is the same result as the nonrandom case. 
        We confirm the irrelevance of $\sigma_\gamma$ numerically in Fig.~\ref{fig:sig_gamma}.

        We therefore ignore the effects of randomness in $\gamma$ completely in our simulations, and focus instead on the effects of charge noise, dissipation, and experimental error in the gate timing.

    \subsection{Numerical Calculations}

     \begin{figure*}[htp] 
        \centering
        \includegraphics[scale=0.43]{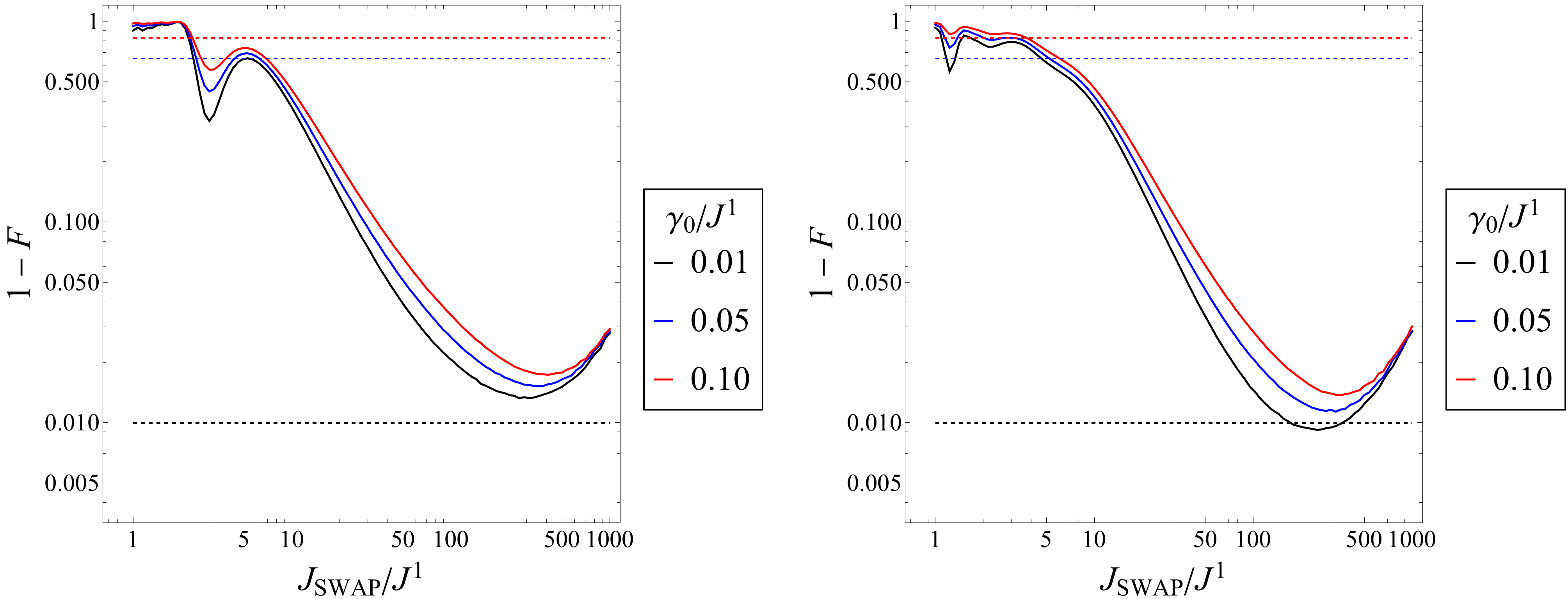}
        \caption{\label{fig:gamma} The infidelity $1-F$ as a function of $J_\text{SWAP}/J^1$ for several values of $\gamma_0$ is shown. 
        For both cases, the spin chain has a length of six. 
        In decreasing order, the dotted lines correspond to the fidelity of the SWAP sequence assuming that the single gate fidelity matches the value achieved by Petta {\it et al}. \cite{Petta2019}, the value achieved by Nichol {\it et al}. \cite{Kandel2019}, and the value necessary for quantum error-correcting codes.
        The plot on the left is for the initial state of $\ket{\Psi_0} = \ket{\uparrow \downarrow \downarrow \downarrow \downarrow \downarrow}$ and the plot on the right is for the initial state of $\ket{\Psi_0} = \ket{S} \otimes \ket{\downarrow \downarrow \downarrow \downarrow}$.
        For both plots, $\sigma_J/J^1 = 0.01$ and $\sigma_\tau = 0.20$ ns.}
    \end{figure*}

        Unlike the effects of dissipation, we must treat the other sources of errors in our SWAP sequence, the charge noise and gate timing error, numerically.  
        We examine two initial states, one being a spin eigenstate with the first spin up and the rest down, $\ket{\Psi_0}=\ket{\uparrow\downarrow\downarrow\downarrow\downarrow\downarrow}$, and the other being an entangled state with the first two spins in a singlet state and the rest being spin down, $\ket{\Psi_0}=\ket{S}\otimes\ket{\downarrow\downarrow\downarrow\downarrow}$, where $\ket{S}=\frac{1}{\sqrt{2}}(\ket{\uparrow\downarrow}-\ket{\downarrow\uparrow}$). 
        In both cases, we transport the spin state of the first qubit to the last qubit and then back again, performing a total of $N = 2(L - 1)$ SWAP gates.  
        In the second case, in which the first two spins are in a singlet state, the first and last SWAP gates, under ideal conditions, would simply yield the same state with an (unimportant for measurement purposes) overall minus sign.
        
    \begin{figure*}[htp] 
        \centering
        \includegraphics[scale=0.43]{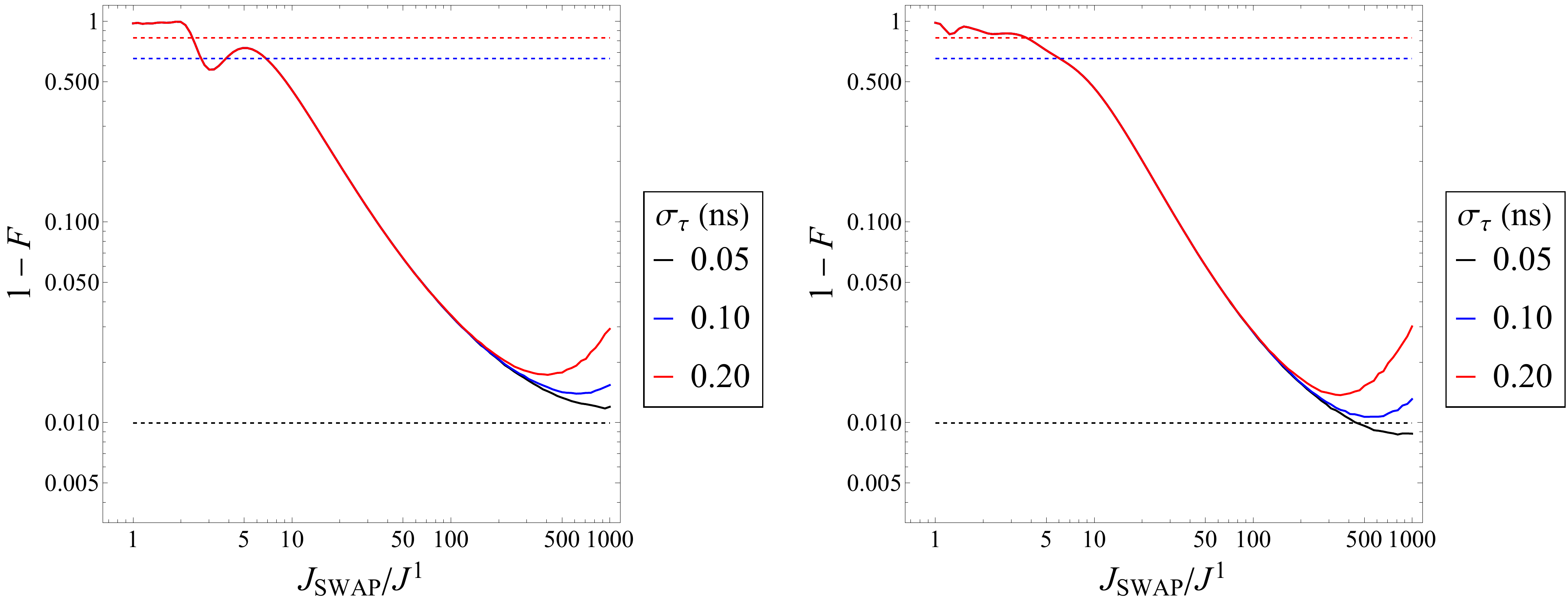}
        \caption{\label{fig:sigtau} The infidelity $1-F$ as a function of $J_\text{SWAP}/J^1$ for several values of $\sigma_\tau$ is shown. 
        For both cases, the spin chain has a length of six. 
        In decreasing order, the dotted lines correspond to the fidelity of the SWAP sequence assuming that the single gate fidelity matches the value achieved by Petta {\it et al}. \cite{Petta2019}, the value achieved by Nichol {\it et al}. \cite{Kandel2019}, and the value necessary for quantum error-correcting codes.
        The plot on the left is for the initial state of $\ket{\Psi_0} = \ket{\uparrow \downarrow \downarrow \downarrow \downarrow \downarrow}$ and the plot on the right is for the initial state of $\ket{\Psi_0} = \ket{S} \otimes \ket{\downarrow \downarrow \downarrow \downarrow}$.
        For both plots, $\sigma_J/J^1 = 0.01$ and $\gamma_0/J^1 = 0.10$.}
    \end{figure*}
    
        For all calculations, we set $\beta = 0.01$ and the length of the spin chain $L = 6$.
        We incorporate charge noise and gate timing error by choosing values from a Gaussian distribution so that $J^k \sim \mathcal{N}(J^k_0,\sigma_J)$ and $\tau \sim \mathcal{N}(\tau_0,\sigma_\tau)$, where $J^k_0$ and $\tau_0$ are the base (noiseless) values. 
        While, in principle, we could have chosen a different distribution for the gate timing error, the exact choice of distribution would only result in quantitative differences from the results that we obtain using the Gaussian distribution. 
        We assume that $\sigma_J$ scales linearly with $J$ \cite{Reed2016,science.1217692} and that $\sigma_\tau$ is a constant value, arising from experimental clock errors in the gate timing.
        Because $\tau_0 \propto J^{-1}_\text{SWAP}$, charge noise also causes errors in the gate duration; the gate timing error $\sigma_\tau$ is an additional source of error in the duration $\tau$, separate from that caused by charge noise.
        For each value of $\sigma_J$ and $\sigma_\tau$, we average the fidelity over 10,000 different realizations of disorder in order to obtain a benchmarked typical fidelity. 
        
    \begin{figure*}[htp] 
        \centering
        \includegraphics[scale=0.43]{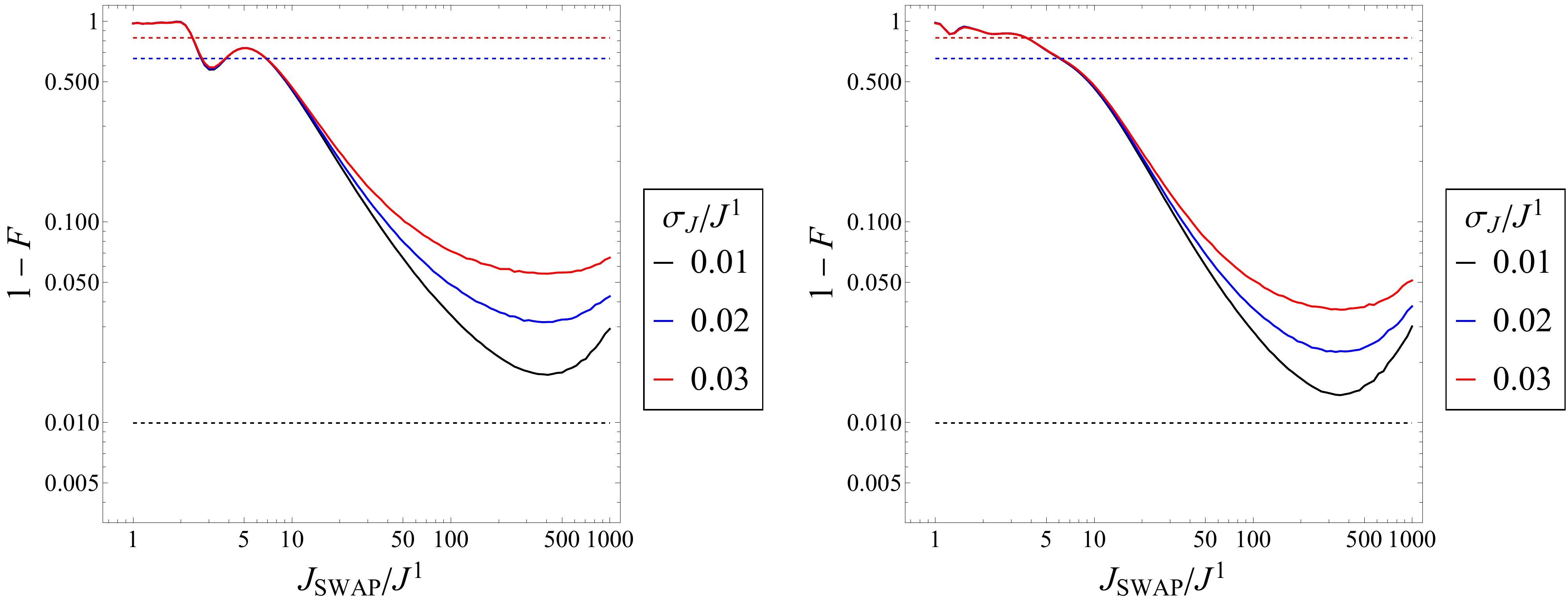}
        \caption{\label{fig:sigJ} The infidelity $1-F$ as a function of $J_\text{SWAP}/J^1$ for several values of $\sigma_J$ is shown. 
        For both cases, the spin chain has a length of six. 
        In decreasing order, the dotted lines correspond to the fidelity of the SWAP sequence assuming that the single gate fidelity matches the value achieved by Petta {\it et al}. \cite{Petta2019}, the value achieved by Nichol {\it et al}. \cite{Kandel2019}, and the value necessary for quantum error-correcting codes.
        The plot on the left is for the initial state of $\ket{\Psi_0} = \ket{\uparrow \downarrow \downarrow \downarrow \downarrow \downarrow}$ and the plot on the right is for the initial state of $\ket{\Psi_0} = \ket{S} \otimes \ket{\downarrow \downarrow \downarrow \downarrow}$.
        For both plots, $\sigma_\tau = 0.20$ ns and $\gamma_0/J^1 = 0.10$.}
    \end{figure*}
    
        The effects of dissipation are seen in Fig.~\ref{fig:gamma}. 
        We see here that as dissipation becomes a more significant factor, SWAP fidelities decrease. 
        As explained in Section~\ref{analytic}, we are able to analytically conclude that the effects of dissipation are completely accounted for by an exponential factor of $e^{-2\gamma_0 t}$, assuming that $\sigma_\gamma$ is small. 
        This simple form implies that as one increases the SWAP coupling strength and decreases the gate duration $\tau$, the effects of dissipation can be overcome.
        This is in fact what our numeric calculations show when excluding the effects of gate timing error---that any effects of the dissipation can be completely overcome by simply increasing $J_\text{SWAP}$.
        
        However, this picture is complicated by the fact that as $\tau$ gets shorter and shorter, the gate timing error dominates, and the infidelity reaches a minimum before increasing. 
        This is seen in Fig.~\ref{fig:sigtau}. 
        Evidently, gate timing error only affects the fidelity for large values of $J_\text{SWAP}$. 
        This makes sense from the point of view that $J_\text{SWAP} \propto \tau^{-1}$, so that as $J_\text{SWAP}$ increases, so does the ratio of $\sigma_\tau/\tau$.
        With the presence of \emph{any} gate timing error, there is an optimal value of $J_\text{SWAP}$ beyond which the fidelity will begin to decrease.
        This minimum infidelity is dependent on all three of our parameters---gate timing error, charge noise, and dissipation.
        
        In Fig.~\ref{fig:sigJ}, we plot the infidelity $1 - F$ as a function of $J_\text{SWAP}/J^1$ for several values of $\sigma_J$. 
        The dissipation and gate timing errors are also included. 
        As charge noise $\sigma_J$ decreases, the infidelities reach smaller and smaller minimum values. 
        Charge noise primarily affects fidelity by affecting $\tau$, the ideal gate duration. 
        Because of charge noise, the experimenter will get the ideal gate duration wrong, since the true value of $J_\text{SWAP}$ is not known.
        Because $\sigma_J \propto J$, increasing $J_\text{SWAP}$ further does not increase fidelity.
        Beyond a certain value of $J_\text{SWAP}$, the infidelity begins to increase; this is again due to gate timing error.
        
        We note that we obtain slightly higher fidelities for large values of $J_\text{SWAP}$ for the singlet state than for the eigenstate case, even though our sequence of SWAP gates remains the same for both cases.  
        However, we see no major qualitative differences between the two cases.
        
    \section{Summary and Conclusion}
        
        We examined the effects of dissipation and gate timing errors on the fidelity of a sequence of SWAP gates on a chain of interacting qubits. 
        We included crosstalk between qubits as well as noise in the interqubit interaction $J^k_n$. 
        We extended previous work by implementing exchange interactions among all qubits, rather than assuming only nearest-neighbor and next-nearest-neighbor interactions.
        We did this by assuming that the overlap of the two wavefunctions decreases exponentially with distance, such that $J^k = \beta^{k-1} J^1$.
        We also included new physics by implementing a dissipation term which may serve as a phenomenological model for valley leakage in Si-based spin qubits and by considering the effects of gate timing error associated with gate operations.

        The effects of dissipation on the fidelity were completely accounted for in our model by a simple exponential factor of $e^{-2 \gamma_0 t}$ for experimentally relevant parameter values ($\sigma_\gamma/\gamma_0 < 0.1$ and $\tau \ll 1 $).
        Therefore, we ignored the effects of randomness in the dissipation in our simulation, and instead focused on how the magnitude of the dissipation parameter itself affects the fidelity.
        The irrelevance of any randomness in the dissipation to quantum computing fidelity is an important finding since it is essentially impossible to control small randomness in dissipation (e.g. valley leakage for Si qubits) arising from uncontrolled (and often, unknown) dissipative mechanisms.

        We also considered the effects of gate timing error $\sigma_\tau$ on SWAP fidelity by selecting $\tau$ from a Gaussian distribution centered on $\tau_0 = \frac{\pi}{4 J_\text{SWAP}}$ with deviation $\sigma_\tau$.
        $J_\text{SWAP}$ in this case refers to the expected value of $J_\text{SWAP}$ rather than the true value after accounting for $\sigma_J$.
        Gate timing error constitutes an additional error source on the gate duration $\tau$ beyond that caused by charge noise.

        We numerically calculated the fidelity of a sequence of SWAP gates that transports a spin state from the left side of a six-qubit chain to the right side, and back again.
        We considered two cases for initial conditions. 
        In the first case, where  $\ket{\Psi_0} = \ket{\uparrow\downarrow\downarrow\downarrow\downarrow\downarrow}$, we ``transported'' the first spin to the sixth and then back to the first. 
        In the second case, where $\ket{\Psi_0} = \ket{S}\otimes\ket{\downarrow\downarrow\downarrow\downarrow}$,  we ``transported'' the entanglement from the second qubit to the sixth and then back to the second.  
        In reality, we started and ended the sequence with a SWAP gate on the first and second qubits as we did in the first case in order to make a more even comparison of the two cases, but each of these gates, under ideal conditions, would yield the same state with an overall minus sign.  
        We examined how the fidelity of the SWAP sequence is affected as we varied parameters corresponding to charge noise, dissipation, and gate timing error.

        We reach several conclusions from our results. 
        First, we confirm that charge noise has an extremely detrimental effect on gate fidelities, and that in order for SWAP gates using Heisenberg exchange to be reliable enough to implement error-correcting codes, our model suggests that relative charge noise ($\sigma_J/J^1$) would have to be less than 0.01, which is a challenge for semiconductor spin qubits, but perhaps not impossible.
        
        Second, the presence of any gate timing error causes a minimum in the infidelity at some value of $J_\text{SWAP}/J^1$. 
        This value is an optimal coupling strength, beyond which infidelity begins to increase. 
        This optimal coupling strength changes with $\sigma_\tau$ and $\gamma_0$, but it is always present.
        Therefore, any hope of completely overcoming dissipation by simply making the exchange coupling stronger and the gate duration shorter should be considered with skepticism, as the two errors are in some sense complementary.
        
        Third, dissipation affects the fidelity of the SWAP operations as expected, with fidelity decreasing as $\gamma_0$ increases. 
        However, we do not expect $\gamma_0/J^1$ to be much greater than $0.1$, and our results seem to imply that even with the leakage parameter saturated at this experimental maximum, the fidelity is not far off from the fidelity achieved with a much lower value of $\gamma_0$.
        Thus, a possible outcome for Si spin qubits is that valley leakage may not be a serious problem as long as it is small.
        
        We do not claim that dissipation is not an issue, simply that we do not see the same dramatic improvements in fidelity when turning down dissipation as we do when turning down charge noise.  
        Thus, an important conclusion is that, at the current stage of Si qubit development, charge noise is a much more overwhelming problem than dissipation induced by valley leakage, but in the future, when charge noise is under better control, it is possible for the valley leakage problem to detrimentally affect the SWAP gate fidelity in a significant way.
        
        The issues that we studied here are not yet of immediate urgency to current experimental Si-based spin qubit systems, where measurements have just started studying SWAP and multiqubit operations, simply because Si-based spin qubits are far behind in development compared with other competing quantum computing platforms such as superconducting circuits and ion traps.  
        However, the fundamental issues discussed theoretically in this work will be crucial in any future development of Si-based spin quantum computing architectures as neither gate timing error nor dissipation can ever be completely suppressed.
        
    \begin{acknowledgments}

        The authors acknowledge the University of Maryland supercomputing resources (http://hpcc.umd.edu) made available for conducting the research reported in this paper.
        This work is supported by the Laboratory for Physical Sciences. 

    \end{acknowledgments}
    
    \bibliography{references.bib}   

\end{document}